\documentclass[12pt]{article}
\usepackage{amsmath}
\usepackage[dvips]{graphicx}
\usepackage{epsfig}
\usepackage{amsmath}
\usepackage{amssymb}
\usepackage{graphics,epstopdf}
\usepackage{amsfonts}
\usepackage{amsthm}
\usepackage[usenames]{color}

\definecolor{AV}{rgb}{0.65,0.0,0}
\definecolor{GC}{rgb}{0,0.0,0.65}
\definecolor{WS}{rgb}{0,0.65,0}

\usepackage[
      colorlinks=true,
      linkcolor=blue,
      urlcolor=blue,
      filecolor=blue,
      citecolor=blue,
      pdfstartview=FitV,
      pdftitle={},
      pdfauthor={},
      pdfsubject={},
      pdfkeywords={},
      pdfpagemode=None,
      bookmarksopen=true
]{hyperref}

\newcommand{\bm}{\begin{multiline}}
\newcommand{\beq}{\begin{equation}}
\newcommand{\eeq}{\end{equation}}
\newcommand{\beqs}{\begin{eqnarray}}
\newcommand{\eeqs}{\end{eqnarray}}

\newcommand{\ra}{\rightarrow}

\begin{document}

\thispagestyle{empty}

\hfill{}

\hfill{}

\hfill{}

\vspace{32pt}

\begin{center}

\textbf{\Large Charged black holes with dark halos}

\vspace{48pt}

\textbf{ Cristian Stelea,}\footnote{Corresponding author e-mail: \texttt{cristian.stelea@uaic.ro}}
\textbf{Marina-Aura Dariescu,}\footnote{E-mail: \texttt{marina@uaic.ro}}
\textbf{Ciprian Dariescu, }\footnote{E-mail: \texttt{ciprian.dariescu@uaic.ro}}

\vspace*{0.2cm}

\textit{$^1$ Department of Exact and Natural Sciences, Institute of Interdisciplinary Research,}\\[0pt]
\textit{``Alexandru Ioan Cuza" University of Iasi}\\[0pt]
\textit{11 Bd. Carol I, Iasi, 700506, Romania}\\[.5em]

\textit{$^{2,3}$ Faculty of Physics, ``Alexandru Ioan Cuza" University of Iasi}\\[0pt]
\textit{11 Bd. Carol I, Iasi, 700506, Romania}\\[.5em]

\end{center}

\vspace{30pt}

\begin{abstract}
Recently, Cardoso et al. \cite{Cardoso:2021wlq} constructed an exact solution of Einstein's equations that describes a supermassive black hole immersed into a dark matter halo. In this work use a solution-generating technique, which is a direct generalization of some of the Ehlers-Harrison transformation in the Ehlers formalism in order to construct in an exact analytical form the charged version of the Cardoso et al. solution. We describe some of its physical properties and, finally, we also present its magnetized version.

\end{abstract}

\vspace{32pt}

\setcounter{footnote}{0}

\newpage

\section{Introduction}

The recent advances in the gravitational-wave astronomy \cite{LIGOScientific:2016aoc} - \cite{LIGOScientific:2021djp} and of the very long baseline interferometry \cite{EventHorizonTelescope:2019dse} allowed us to shed light onto an exciting new chapter of the physics of black holes and other compact objects, which was so far beyond our experimental reach (see for instance \cite{Barack:2018yly}). In this new gravitational-wave astronomy era one has new opportunities to test General Relativity with unprecedented levels of precision, to impose new experimental constraints on GR and even to exclude possible modifications of it.

As end points of gravitational collapse of massive stars, black holes play a central role in General Relativity (GR) and their properties have been studied extensively in the recent decades. One of their characteristic features is their uniqueness. Basically this means that all electrovacuum black-holes in four dimensions are characterized by their mass, angular momentum and electric charge, or, in other words, they belong to the class of Kerr-Newman black holes (for a recent  review of the no-hair theorems see \cite{Chrusciel:2012jk} and references therein). One should note at this point that the no-hair theorems apply to isolated, stationary and regular black holes in the Einstein-Maxwell theory. However, it is very unlikely that black holes are isolated objects in the universe. Instead, one actually expects that they live in a very complex environment composed of charged plasma, electromagnetic fields and dark matter. Nonetheless, the no-hair theorem can still be tested for supermassive black holes in the center of the galaxies \cite{Cardoso:2016ryw}.

While there is compelling evidence at all astrophysical scales for the existence of dark matter in galaxies (see for instance  \cite{Clowe:2006eq}), the nature of the dark matter remains so far elusive \cite{Bertone:2016nfn}, as the standard model of particle physics does not contain any suitable particles that are able to explain this component of the matter in universe \cite{Bertone:2004pz}. Since dark matter is assumed to interact only gravitationally, the detection of gravitational waves has opened up new opportunities to explore the physics of dark matter \cite{Bertone:2019irm}, \cite{Barack:2018yly}. Indeed, it is known that dark matter may cluster at the center of the galaxies, close to the supermassive black hole \cite{Sadeghian:2013laa}. As such, the dark matter halo around the black holes is expected to leave detectable imprints on the dynamics and the emitted gravitational wave signals from such galactic systems \cite{Barausse:2014tra}. For a better understanding of how the galactic dark matter distribution affects the dynamics of coalescing binaries of compact objects (formed from black holes and/or neutron stars) and the gravitational waves emitted in this process, one needs a fully relativistic solution of Einstein's equations that would describe a black hole immersed in such a medium.

Remarkably, Cardoso et al. \cite{Cardoso:2021wlq} found an exact solution that describes a black hole immersed into a galactic distribution of matter. This solution may serve as a model for a supermassive black hole in the center of a galaxy surrounded by a dark matter halo. There are various profiles of matter distribution in galactic halos that have been used to model black holes in realistic dense environments and their choice is usually dictated by the size, mass and the form of the galaxy \cite{Navarro:1995iw} - \cite{Taylor:2002zd}. The exact solution of the Cardoso et al. was based on a Hernquist-type density distribution \cite{Hernquist1}. This profile is well suited to describe the galactic nuclei observed in elliptical galaxies or the bulges of spiral galaxies. Other black holes models with different dark matter halo distributions have been obtained in \cite{Konoplya:2022hbl} and \cite{Jusufi:2022jxu}. In \cite{Figueiredo:2023gas} it was further developed a numerical approach to find black holes solutions embedded in an anisotropic fluid with a generic density profile. 

In this study we construct and study an exact solution of the Einstein's equations that describes the charged version of the Cardoso et al. solution. This solution is constructed by using a previously known solution generating technique developed specifically for interior fluid solutions \cite{Stelea:2018shx} and in a more general form in \cite{Stelea:2018elx} (see also \cite{Yazadjiev:2004bg}). An electrically charged generalization of the Cardoso et al. solution has been presented in \cite{Feng:2022evy}. However, their solution has been studied only numerically. By contrast, the electrically charged solution that we shall present in our work is an exact solution of the Einstein's equations that directly generalizes the Cardoso et al. solution in presence of an electric field. 

This work is organized as follows. In Section $2$ we review the solution found in \cite{Cardoso:2021wlq}. In Section $3$ we present the solution generating method and use it to construct in exact analytical form the electric version of the Cardoso et. al. solution. In Section $4$ we address some of its physical properties and compare it with the numerical solutions proposed in \cite{Feng:2022evy}. Section $5$ is dedicated to conclusions and avenues for further work. 

In what follows we use geometric units such that $c=G=1$.

\section{Black hole surrounded by the galactic halo}

In order to embed a black hole into a dark matter halo, the authors of \cite{Cardoso:2021wlq} proposed a method based on a generalized version of an Einstein cluster (see for instance \cite{Ruffini}). Basically, in an Einstein cluster the particles move freely under the influence of the gravitational field produced by all of them together. Since in the cluster the particles do not collide, there is no radial pressure and the matter distribution is modeled by using an anisotropic fluid with vanishing radial pressure \cite{Lake:2006pp}.

To describe the matter distribution that is observed in elliptical galaxies and the bulges of the spiral galaxies, Hernquist proposed the following density distribution \cite{Hernquist1} :
\beqs
\rho&=&\frac{Ma_0}{2\pi r(r+a_0)^3},
\label{rhenq}
\eeqs
where $M$ is the mass of the galactic halo and $a_0$ is a typical lengthscale. Note that for astrophysical applications of this density profile one should require $M/a_0\sim 10^{-4}$ \cite{Hernquist2}. One should expect that this profile is modified in presence of a supermassive black hole at the center of the galaxy since the dark matter may cluster close to the black hole \cite{Sadeghian:2013laa}.

To address this problem the authors of \cite{Cardoso:2021wlq} used a construction similar to an Einstein cluster. One starts with a spherically symmetric configuration in which the galactic matter is now modelled by an anisotropic fluid with the stress energy of the form:
\beqs
T^{0\mu}_{~~\nu}&=&diag\big[-\rho^0, 0, p_t^0, p_t^0\big],
\label{Tin}
\eeqs
where $\rho^0$ is the energy density of the galactic halo and $p_t^0$ is the tangential pressure \cite{Lake:2006pp}. The geometry is given by:
\beqs
ds^2&=&g_{\mu\nu}^0dx^{\mu}dx^{\nu}=-f(r)dt^2+\frac{dr^2}{1-\frac{2m(r)}{r}}+r^2(d\theta^2+\sin^2\theta d\varphi^2).
\label{min}
\eeqs

For a solution of the Einstein's equations $G_{\mu\nu}^0=8\pi T^0_{\mu\nu}$ one has to impose in the geometry the condition $G^0_{r r}=0$ (the radial pressure vanishes in this system) such that:
\beqs
\frac{f'}{f}&=&\frac{2m(r)/r}{r-2m(r)}.
\label{fp}
\eeqs
One also obtains $\rho^0=\frac{m'}{4\pi r}$, while the tangential pressure can also be obtained from the Bianchi identities as:
\beqs
p^0_t&=&\frac{m(r)}{r-2m(r)}\frac{\rho^0}{2}.
\eeqs
Note that given a mass function $m(r)$ one can compute the remaining quantities of interest $f(r)$, $\rho^0$ and $p_t^0$. The choice in \cite{Cardoso:2021wlq} for the mass function was the following:
\beqs
m(r)&=&M_{bh}+\frac{Mr^2}{(r+a_0)^2}\left(1-\frac{2M_{bh}}{r}\right)^2,
\eeqs
Then, using(\ref{fp}) one can find the following solution such that $f(r\ra\infty)=1$:
\beqs
f(r)&=&\left(1-\frac{2M_{bh}}{r}\right)e^\Gamma,
\eeqs
where $\Gamma=-\pi\sqrt{\frac{M}{\xi}}+2\sqrt{\frac{M}{\xi}}\arctan\frac{r+a_0-M}{\sqrt{M\xi}}$ while $\xi=2a_0-M+4M_{bh}$. The energy density becomes:
\beqs
\rho^0&=&\frac{M(a_0+2M_{bh})\left(1-\frac{2M_{bh}}{r}\right)}{2\pi r(r+a_0)^3}.
\label{rho0}
\eeqs
Note that asymptotically it approaches the Hernquist profile (\ref{rhenq}). For astrophysical systems of interest one has $M/a_0\sim 10^{-4}$ and we shall also assume that the mass of the halo $M$ is much bigger than the mass of the black hole $M_{bh}$ such that $M_{bh}\ll M\ll a_0$. This way one can also assure that $\xi>0$ such that there are no curvature singularities in the geometry except the one in origin \cite{Cardoso:2021wlq}.

Note that $r=2M_{bh}$ is the location of a black hole horizon in the geometry (\ref{min}) as $f(r)$ vanishes there, while the surface $r=2m(r)$ becomes a null surface. Also, the energy density (\ref{rho0}) vanishes on the horizon while the tangential pressure becomes:
\beqs
p_t^0(r=2M_{bh})&=&\frac{M}{16\pi M_{bh}(2M_{bh}+a_0)^2}.
\eeqs
This means that the weak energy condition (WEC) and the strong energy condition (SEC) are satisfied near the black hole horizon. However the dominant energy condition (DEC) is violated since $\rho^0<p_t^0$ near the horizon. Nonetheless, since the anisotropic fluid has arbitrarily small pressure and density near the horizon, the violation of DEC may play no role in the spacetime dynamics \cite{Cardoso:2021wlq}.

\section{The charged black hole with a dark halo solution}

We shall follow the solution-generating technique described in  \cite{Stelea:2018shx}, \cite{Stelea:2018elx} (see also \cite{Yazadjiev:2004bg}). We start with the initial uncharged geometry (\ref{min}) which is sourced by an anisotropic fluid with the stress-energy tensor given by (\ref{Tin}). Let us define the quantity $\Lambda=\frac{1-U^2f(r)}{1-U^2}$, where $0\leq U<1$ is a constant parameter.

Let us construct now the metric:
\beqs
ds^2&=&g_{\mu\nu}dx^{\mu}dx^{\nu}=-\frac{f(r)}{\Lambda^2}dt^2+\Lambda^2\bigg[\frac{dr^2}{1-\frac{2m(r)}{r}}+r^2(d\theta^2+\sin^2\theta d\varphi^2)\bigg],
\label{mfin}
\eeqs
and use the following ansatz for the electric field:
\beqs
A_{\mu}&=&\left(\frac{Uf(r)}{\Lambda}, 0, 0, 0\right).
\label{potel}
\eeqs
Then the system of Einstein-Maxwell-fluid equations of motion (see also \cite{Feng:2022evy}):
\beqs
G_{\mu\nu}&=&8\pi T_{\mu\nu}+8\pi T_{\mu\nu}^{em},\nonumber\\
F^{\mu\nu}_{~~;\nu}&=&4\pi J^{\mu},~~~~J^{\mu}_{~;\mu}=0,~~~T^{\mu\nu}_{~~;\nu}=F^{\mu\nu}J_{\nu}
\eeqs
is satisfied if one considers the stress-energy of the fluid as being given by:
\beqs
T_{\mu\nu}&=&(\rho+\rho_e+p_t)u_{\mu}u_{\nu}+p_t(g_{\mu\nu}-\chi_{\mu}\chi_{\nu}),
\label{Tfluid}
\eeqs
while the stress-energy tensor of the electromagnetic field can be written in the usual form:
\beqs
T^{em}_{\mu\nu}&=&\frac{1}{4\pi}\left(F_{\mu\gamma}F_{\nu}^{~\gamma}-\frac{1}{4}g_{\mu\nu}F_{\gamma\delta}F^{\gamma\delta}\right).
\eeqs
Here $u^{\mu}=\frac{\Lambda}{\sqrt{f(r)}}\delta^{\mu}_t$ is the $4$-velocity of the fluid, normalized such that $u^{\mu}u_{\mu}=-1$, while $\chi^{\mu}=\frac{\sqrt{1-\frac{2m(r)}{r}}}{\Lambda}\delta^{\mu}_r$ is the unit vector in the radial direction. Also, we defined $\rho=\rho^0/\Lambda^2$ and $p_t=p_t^0/\Lambda^2$, while:
\beqs
\rho_e&=&\frac{2(\rho+2p_t)}{1-U^2}\frac{U^2f(r)}{\Lambda}.
\label{rhoe}
\eeqs
The $4$-current $J_{\mu}=j_t\delta^t_{\mu}$ has the form:
\beqs
j_t&=&-\frac{2(\rho+2p_t)}{1-U^2}\frac{Uf(r)}{\Lambda^2}.
\eeqs
From this expression of the $4$-current one can easily determine the proper charge density:
\beqs
\sigma_e&=&-J_{\mu}u^{\mu}=\frac{2U\sqrt{f(r)}}{\Lambda}(\rho+2p_t).
\label{chargedensity}
\eeqs
Finally, let us note that $\rho_e=J_{\mu}A^{\mu}$ corresponds to the effective stress-energy tensor generated by the source term $\sim J_{\mu}A^{\mu}$ in the Maxwell-fluid Lagrangean, however, we have chosen here to include its contribution in the stress-tensor energy $T_{\mu\nu}$ that describes the anisotropic fluid. With this in mind, we shall use $\rho_f=\rho+\rho_e$ as the total energy density of the charged fluid.

Note that the temporal component of the metric is in this case $g_{tt}=-\frac{f(r)}{\Lambda^2}$ and one can then express it as a function of the electric potential $\phi=\frac{Uf(r)}{\Lambda}$ in the form $g_{tt}=-\phi^2-\frac{1-U^2}{U}\phi$. This belongs to the so-called Weyl-type relations, which was initially introduced by Weyl in his study of static Einstein-Maxwell systems \cite{Weyl:1917gp}. Basically, by assuming a functional dependence of the $tt$ component of the metric Weyl was able to generate an exact solution of the static Einstein-Maxwell equations and showed that the functional dependence must be quadratic for vacuum Einstein-Maxwell systems. This matter was further investigated by Gautreau and Hoffman in \cite{GH}. They considered the structure of sources that could produce such Weyl-type fields and they further showed that the form of the charge density must take precisely the form given in (\ref{chargedensity}) (see also \cite{Lemos:2009mr}). Therefore, since in our solution-generating technique we are producing fields of the Weyl-type then the form of the charge density is fixed by (\ref{chargedensity}). \footnote{Note that for charged dust solutions with $p_t=0$ the charge density approaches the charge density given in equation $(15)$ in the Debye model considered in \cite{Feng:2022evy}.}

One could of course envisage more general types of fields with various other charge densities as in \cite{Feng:2022evy}, however the obtained solutions are so far numerical.

\section{Properties of the charged solution}

First of all, let us notice that the fields defined in the previous section reduce to the Cardoso et al. solution \cite{Cardoso:2021wlq} if one takes $U=0$, as expected. In absence of the galactic halo, if one takes $M=0$ and $a_0=0$ then the charged metric (\ref{mfin}) reduces to the Reissner-Nordstr\"{o}m solution after performing the coordinate transformation $R=r\Lambda$, that is:
\beqs
r&=&R-\frac{2M_{bh}U^2}{1-U^2}
\label{rR}
\eeqs
Indeed, in our coordinates the charged exterior geometry corresponds to:
\beqs
ds^2&=&-\frac{1-\frac{2M_{bh}}{r}}{\Lambda^2}+\frac{\Lambda^2}{1-\frac{2M_{bh}}{r}}dr^2+\Lambda^2r^2(d\theta^2+\sin^2\theta d\varphi^2),
\label{elSchw}
\eeqs
where $\Lambda=\frac{1-U^2\left(1-\frac{2M_{bh}}{r}\right)}{1-U^2}$, while the electromagnetic potential has the only non-zero component:
 \beqs
A_t=\frac{U\left(1-\frac{2M_{bh}}{r}\right)}{\Lambda}.
\eeqs
After performing the coordinate transformation $R=\Lambda r$ as in (\ref{rR}) one recovers the Reissner-Nordstr\"{o}m solution:
\beqs
ds^2&=&-\left(1-\frac{2M_{ADM}}{R}+\frac{Q_{\infty}^2}{R^2}\right)dt^2+\frac{dr^2}{1-\frac{2M_{ADM}}{R}+\frac{Q_{\infty}^2}{R^2}}+R^2(d\theta^2+\sin^2\theta d\varphi^2),\nonumber\\
A_t&=&U+\frac{Q_{\infty}}{R}.
\eeqs

If the initial black hole solution has the mass $M_{bh}$ then the charged black hole solution will have the ADM mass $M_{ADM}$ and the electric charge $Q_{\infty}$ given by:
\beqs
M_{ADM}&=&\frac{1+U^2}{1-U^2}M_{bh},~~~Q_{\infty}=\frac{2M_{bh}U}{1-U^2}.
\eeqs

Let us notice that in the charged geometry (\ref{mfin}) the surface $r=2M_{bh}$ corresponds again to a black hole horizon, namely the horizon of the charged black hole in the center of the dark halo. Near the black hole horizon one has $\Lambda(r\ra2M_{bh})=1/(1-U^2)$. Therefore, the energy density vanishes on the black hole horizon $\rho_f(r\ra2M_{bh})=0$, while the transversal pressure $p_t=p_t^0/\Lambda^2$ does not. This means that in the generated charged solution the WEC and SEC are satisfied while the DEC is still violated, as it happens in the uncharged case.
\begin{figure}[ht!]
\begin{center} 
\includegraphics[height=.34\textwidth, angle =0 ]{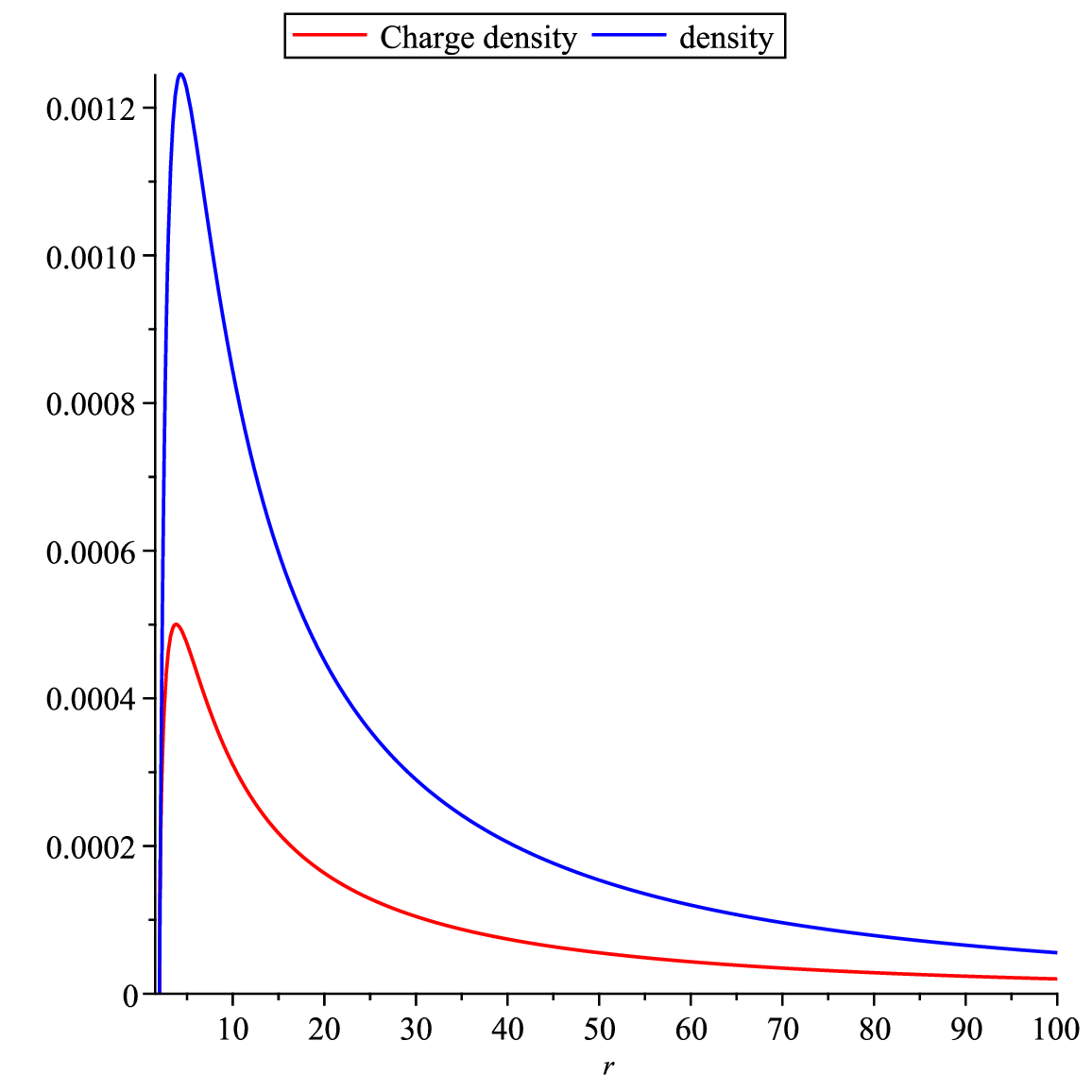}
\includegraphics[height=.34\textwidth, angle =0 ]{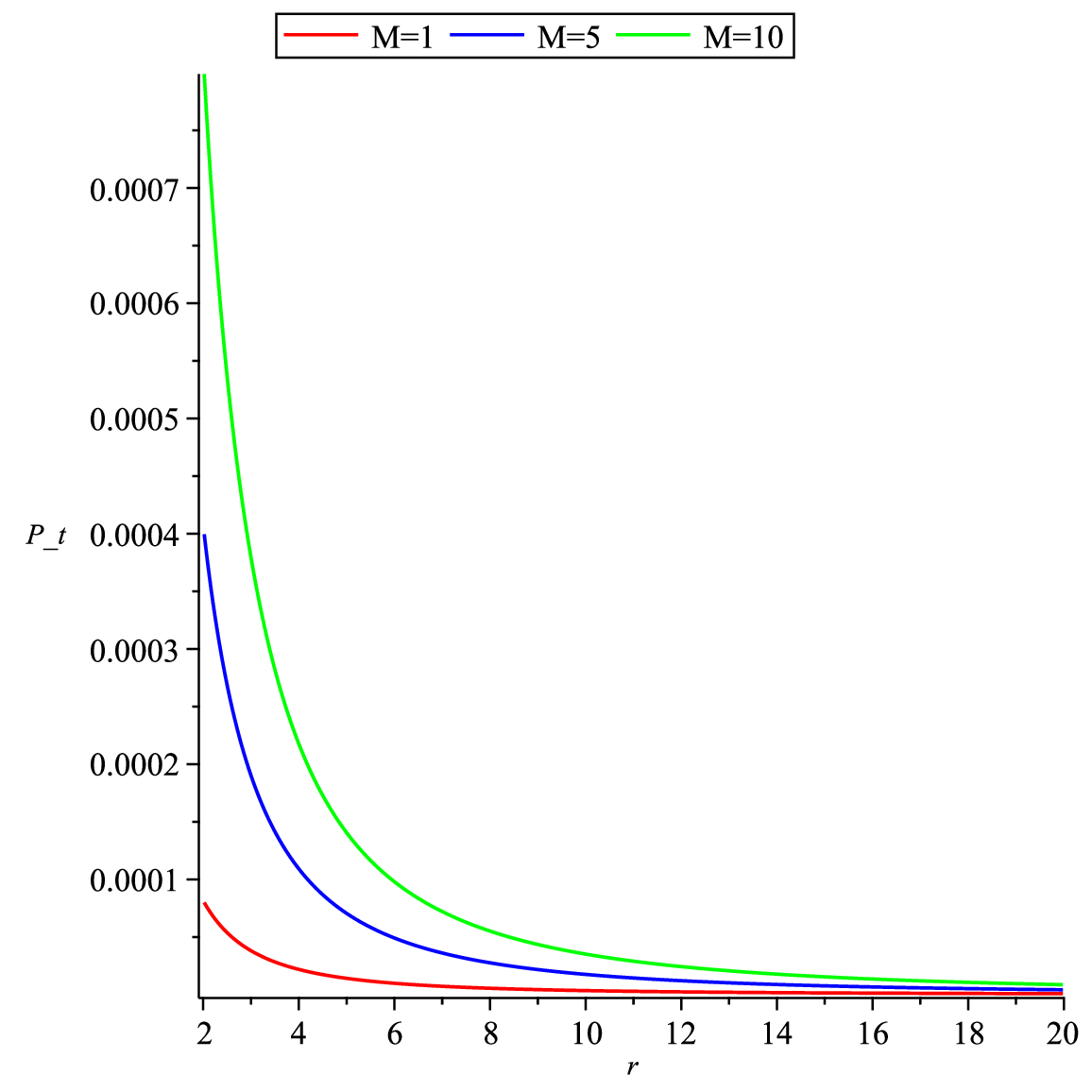}
\end{center}
\caption{
 {\it Left panel}:
The total density $\rho_f$  and the charge density $\sigma_e$ for $M_{bh}=1$, $a_0=200$ and $M=10$, while $U=0.3$.
 {\it Right panel}:
The tangential pressure $p_t$ for $M_{bh}=1$, $a_0=200$ and $U=0.3$ for various values of the halo mass $M=1$, $M=5$ and $M=10$.
}
\label{rho-p}
\end{figure}

In Figure \ref{rho-p} we plotted the total density $\rho_f$, the charged density $\sigma_e$ and the transversal pressure $p_t$ for various values of the parameters.

Turning now our attention to the asymptotic region, let us notice that asymptotically $\Lambda\ra 1$ as $r\ra\infty$ and the charged geometry (\ref{mfin}) is asymptotically flat, as expected. While at infinity one has $f(r)\ra 1-\frac{2(M_{bh}+M)}{r}$, after performing the coordinate transformation (\ref{rR}) to canonically normalize the $2$-sphere factor in the charged geometry (\ref{mfin}), one obtains then the ADM mass of the charged geometry as being given by:
\beqs
M_{ADM}&=&\frac{1+U^2}{1-U^2}(M_{bh}+M).
\eeqs
It contains a contribution from both the black hole mass $M_{bh}$ and also from the dark halo mass, $M$. One can also expect from these considerations that the electric charge of the charged black hole with the dark halo, as measured at infinity should be given by:
\beqs
Q_{\infty}&=&\frac{2U(M_{bh}+M)}{1-U^2}.
\label{qi}
\eeqs

Let us compute now the electric field $E^{\mu}$ for our charged geometry. In general one has:
\beqs
E^{\mu}&=&F^{\mu\nu}u_{\nu}=\frac{1}{\Lambda}\frac{\Phi'}{\sqrt{f(r)}}\left(1-\frac{2m(r)}{r}\right)\delta^{\mu}_r.
\label{elfield}
\eeqs
Here we denoted by $\Phi=\frac{Uf(r)}{\Lambda}$ the electric potential defined in (\ref{potel}). 
\begin{figure}[ht!]
\begin{center} 
\includegraphics[height=.34\textwidth, angle =0 ]{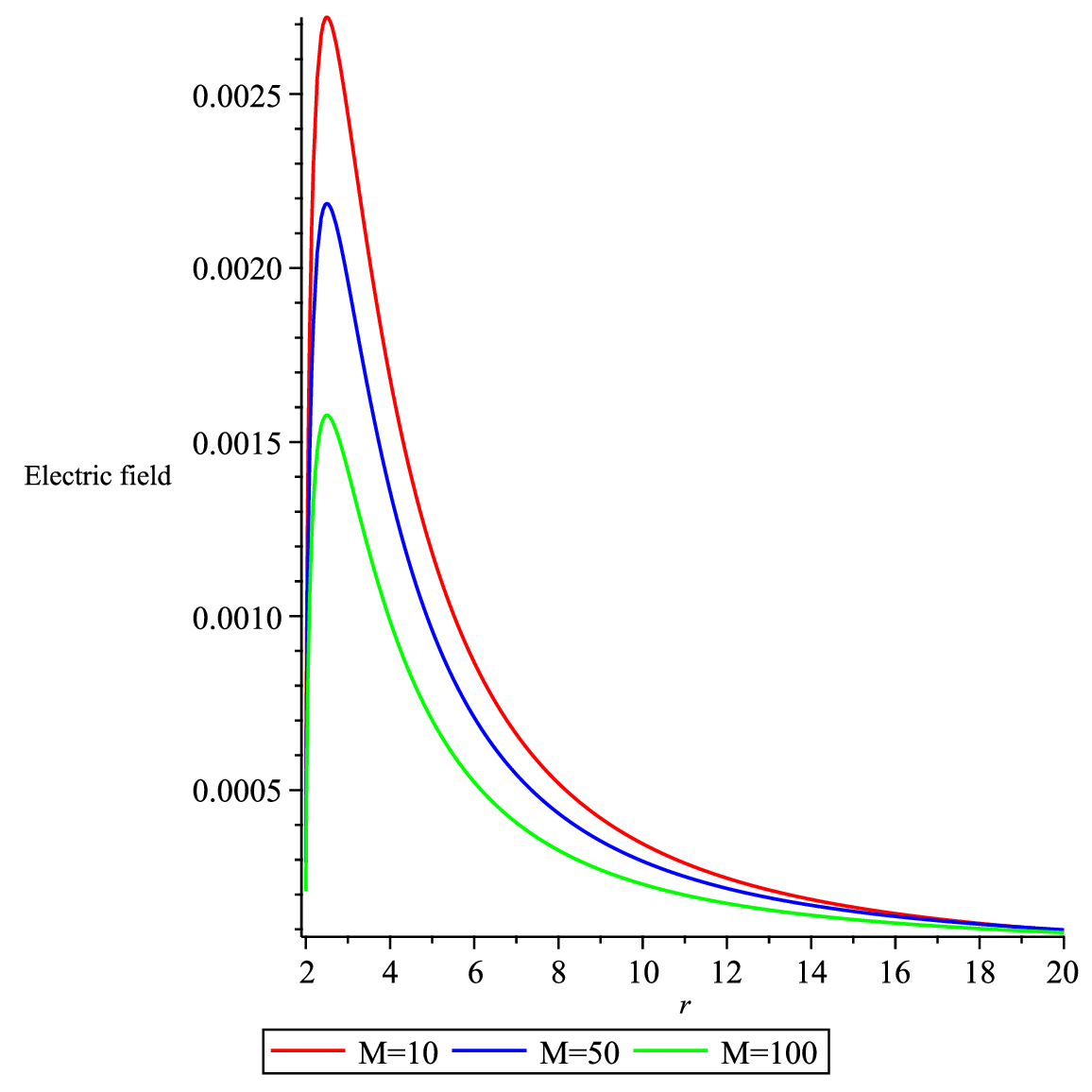}
\includegraphics[height=.34\textwidth, angle =0 ]{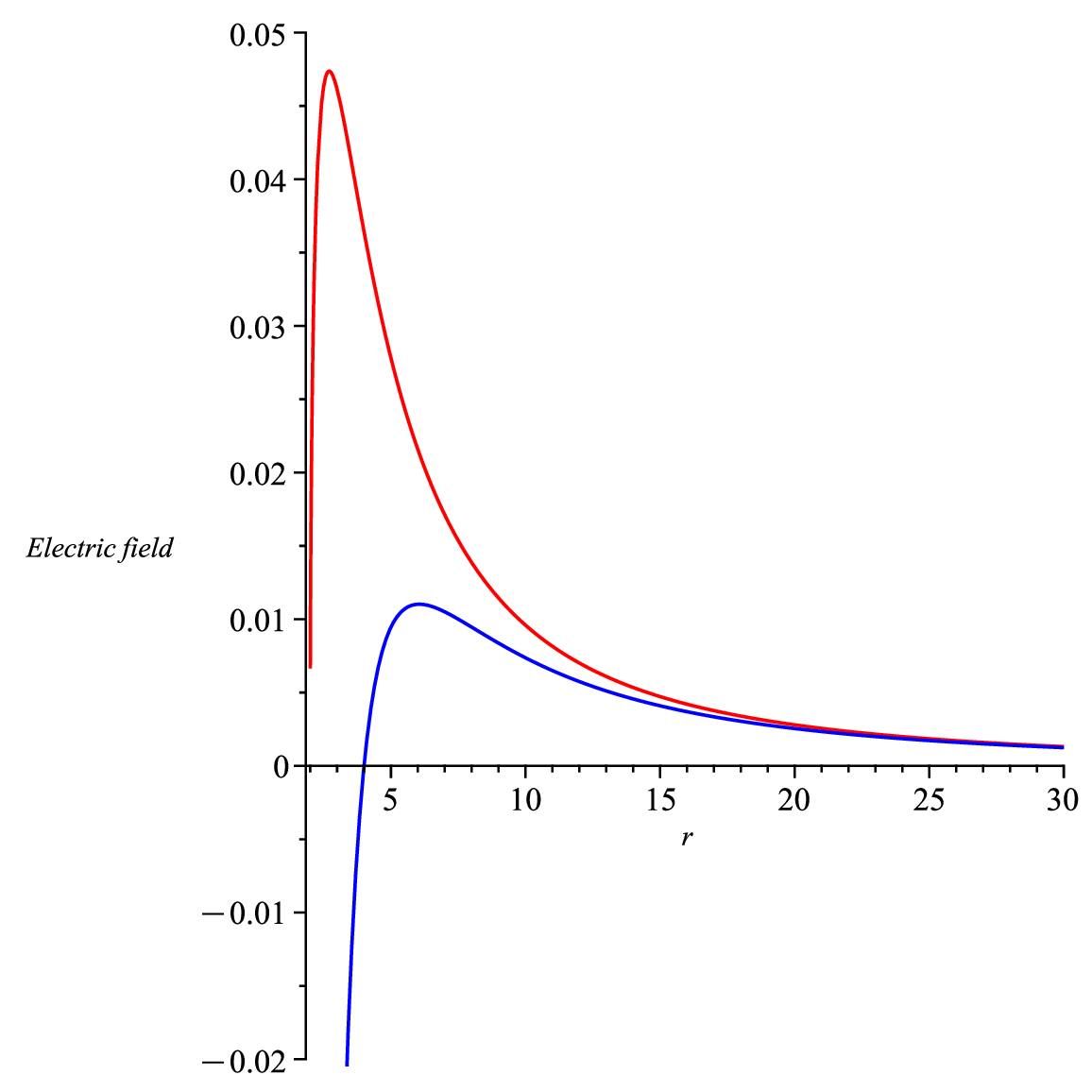}
\end{center}
\caption{
 {\it Left panel}:
The electric field defined in (\ref{elfield}) for $M_{bh}=1$, $a_0=200$ and various values of the halo's mass $M$, while $U=0.02$.
 {\it Right panel}:
The electric field (in red) from (\ref{elfield}) for $M_{bh}=1$, $M=10$ and $a_0=1000$ versus the electric field (in blue) generated by a Schwarzschild black hole with mass $M_{bh}=1$ after we apply our charging technique on it. The charging parameter is $U=0.5$.
}
\label{el-f}
\end{figure}
In Figure \ref{el-f} we plotted the radial electric field for various values of the parameters.

One can use this expression of the electric field together with the Gauss theorem to compute the electric charge of the black hole and also the total electric charge as measured at infinity. However, one arrives at the same results by using the usual formula:
\beqs
Q&=&\frac{1}{4\pi}\int_{S^2}\star F,
\label{charge}
\eeqs
where we defined $\star F_{\mu\nu}=\frac{1}{2}\epsilon_{\mu\nu\gamma\delta}F^{\gamma\delta}$ as the Hodge-dual tensor, while $\epsilon_{\mu\nu\gamma\delta}$ is the Levi-Civita tensor. Here $S^2$ is a $2$-sphere on which we compute the charge. In our case one obtains:
\beqs
(\star F)_{\theta\varphi}&=&\frac{\Phi'\Lambda^2}{\sqrt{f(r)}}\sqrt{1-\frac{2m(r)}{r}}r^2\sin\theta.
\eeqs
Turning now our attention in the near-horizon region, note that $m(r\ra2M_{bh})=M_{bh}$, while $\Lambda(r\ra2M_{bh})=\frac{1}{1-U^2}$, while $f'\sim \frac{2M_{bh}}{r^2}e^\Gamma$, with $e^{\Gamma}\sim1-\frac{2M}{a_0}$. Integrating now (\ref{charge}) on the horizon one obtains the black hole charge:
\beqs
Q_{bh}&=&\frac{2M_{bh}U}{1-U^2}e^{\Gamma}.
\eeqs
Note that the presence of the dark matter halo surrounding the black hole manifests itself in the factor $e^{\Gamma}\sim1-\frac{2M}{a_0}$.
Using now the same formula (\ref{charge}) in the asymptotic region one recovers the electric charge at infinity (\ref{qi}).

Let us turn now our attention to the location of the light rings in the charged geometry (\ref{mfin}). Given the fact that the $2$-sphere geometry in (\ref{mfin}) is not canonically normalized, we shall make use of a slightly more general spherical geometry than (\ref{mfin}) to determine the location of the light rings. Consider then the following more general spherically symmetric geometry:
\beqs
ds^2&=&-A(r)dt^2+\frac{dr^2}{B(r)}+C(r)(d\theta^2+\sin^2\theta d\varphi^2).
\eeqs
Due to the spherical symmetry, one can consider the motion on the equatorial plane $\theta=\pi/2$, such that the Langangean describing the geodesics of this geometry takes the following form:
\beqs
{\cal L}&=&\frac{1}{2}\bigg[-A(r)\dot{t}^2+\frac{\dot{r}^2}{B(r)}+C(r)\dot{\varphi}^2\bigg].
\eeqs
Since $t$ and $\varphi$ are cyclic coordinates one can introduce two conserved quantities, namely the energy $E=A(r)\dot{t}$ and the angular momentum $L=C(r)\dot{\varphi}$, as measured at infinity. Then, for null geodesics ${\cal L}=0$ one obtains:
\beqs
\frac{A(r)}{B(r)}\frac{\dot{r}^2}{L^2}+V_{eff}(r)=\frac{1}{b^2},
\eeqs
where $b=\frac{L}{E}$ is the impact parameter, while $V_{eff}(r)=\frac{A(r)}{C(r)}$ is the effective potential for null geodesics. A circular null geodesic at $r=r_c$ will satisfy the conditions $\dot{r_c}=0$ and $\ddot{r}_c=0$, that is $V_{eff}(r_c)=\frac{1}{b_c^2}$ and $V'_{eff}(r_c)=0$. Note that $V_{eff}(r\ra2M_{bh})=V_{eff}(\infty)=0$. If $V''_{eff}(r_c)>0$ one obtains a null stable circular geodesic, which corresponds to an anti-photon light ring. If the circular null orbit is unstable then $V''_{eff}(r_c)<0$ and it corresponds to an usual light ring.

\begin{figure}[ht!]
\begin{center} 
\includegraphics[height=.34\textwidth, angle =0 ]{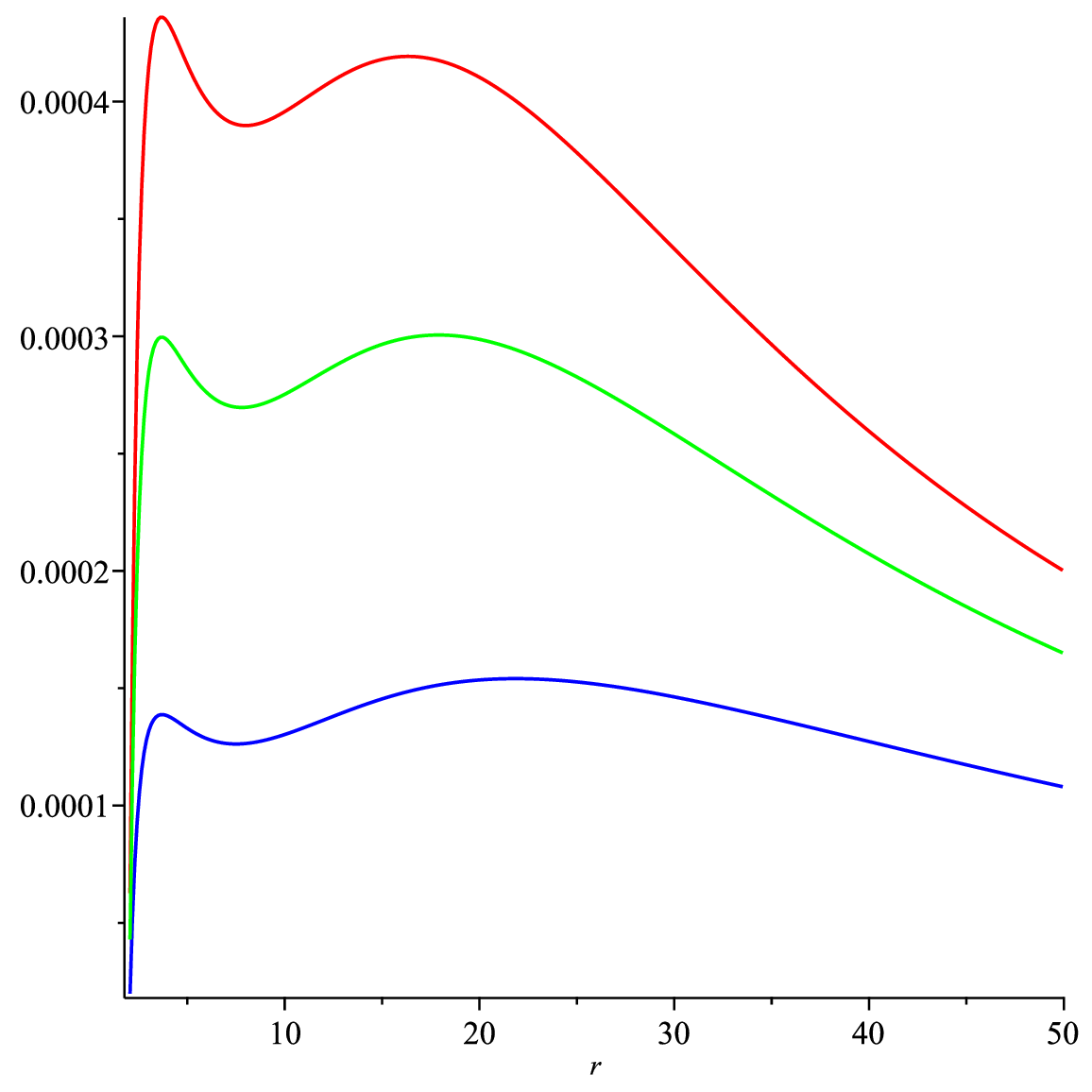}
\end{center}
\caption{
The effective potential $V_{eff}$ for $M_{bh}=1$, $M=15$, $a_0=10$. Here the charging parameter takes the values $U=0$ (for the red curve), $U=0.3$ for the blue curve, while $U=0.5$ for the green one.
}
\label{effpot}
\end{figure}
A necessary condition for the existence of an anti-photon sphere outside a regular black hole horizon is the violation of the DEC or SEC \cite{Cvetic:2016bxi}, \cite{Guo:2022ghl}. In particular, this is what happens in our case as well, as can be seen in Figure \ref{effpot}.

 In the uncharged geometry (\ref{min}) it is well known by now that there are parameter values for which the geometry admits multiple light rings \cite{Guo:2022ghl}, \cite{Xavier:2023exm}. Keeping only terms up to ${\cal O}\left(\frac{1}{a_0^3}\right)$ one obtains \cite{Cardoso:2021wlq}:
\beqs
r_c&=&3M_{bh}\left(1+\frac{MM_{bh}}{a_0^2}\right), ~~~ b_c=3\sqrt{3}M_{bh}\left(1+\frac{M}{a_0}+\frac{5M^2-18MM_{bh}}{6a_0^2}\right).
\eeqs
Now, in the final charged geometry (\ref{mfin}) one has $A(r)=f(r)/\Lambda^2$, $B(r)=\frac{1-\frac{2m(r)}{r}}{\Lambda^2}$ and $C(r)=r^2\Lambda^2$. Expanding up to the terms  ${\cal O}\left(\frac{1}{a_0^2}\right)$ one obtains the light ring location:
\beqs
r_c&=&\frac{3-U^2+\sqrt{9U^4-14U^2+9}}{2(1-U^2)}M_{bh}-\frac{1}{a_0}\frac{2MM_{bh}}{(1-U^2)^2}\left(1+\frac{1+U^2}{\sqrt{9U^4-14U^2+9}}\right),
\eeqs
while the critical impact parameter $b_c$ up to terms ${\cal O}\left(\frac{1}{a_0}\right)$ becomes:
\beqs
b_c&=&\frac{\left(U^2(2M_{bh}-R_0)+R_0\right)^2}{(1-U^2)\sqrt{R_0(R_0-2M_{bh})}},
\eeqs
where $R_0=\frac{3-U^2+\sqrt{9U^4-14U^2+9}}{2(1-U^2)}M_{bh}$.

Note that for small values of the parameter $U$ then the charge becomes $Q\sim2UM$, so it also corresponds to a small charge limit. Then in this small charge limit one has in the charged geometry (\ref{mfin}):
\beqs
r_c&\sim&3M_{bh}+\frac{4}{3}M_{bh}U^2+...=3M_{bh}+\frac{Q^2}{3M_{bh}}+...
\eeqs

This corresponds to the location of an ustable light ring in the charged geometry.

\section{Conclusions}

There is by now compelling evidence for the existence of dark matter, at all astronomical scales. Using the new experimental advances that started a new era in gravitational-wave astronomy, there are promising prospects in using gravitational waves to search for the presence of dark matter around compact objects in galaxies \cite{Bertone:2019irm}. As it is well-known, the dark matter may cluster at the center of the galaxies, close to the supermassive black hole \cite{Sadeghian:2013laa}. To study the influence of the galactic dark matter distribution on the properties of a suppermassive black hole one needs a fully relativistic solution of Einstein's equations that would describe a black hole immersed in such a medium. Remarkably, Cardoso et al. \cite{Cardoso:2021wlq} found an exact solution that describes a black hole immersed into a galactic distribution of matter.

In our work we further generalized this exact solution by including the effects of an electric field. This new exact solution was constructed by applying a previously known solution generating technique. In Section $3$ we constructed the new charged solution while in Section $4$ we presented some of its physical properties. We computed the mass and the charge at infinity for the new solution and the electric field around the black hole. Finally, we computed the location of the light rings in this new solution. For certain values of the parameters (see Figure \ref{effpot}) one can see from the maxima of the effective potential $V_{eff}(r)$ the presence of two light rings and of one anti-photon light ring. If the potential peak at the inner photon sphere is lower than that at the outer photon sphere then the light rays from the vicinity of the first light ring cannot escape at infinity, effectively rendering invisible this photon ring to distant observers.

It is also possible to generalize the Cardoso et al. solution by including a poloidal magnetic field by using the techniques from \cite{Stelea:2018cgm}, \cite{Yazadjiev:2011ks} and \cite{Stelea:2018elx}. If one defines $\Lambda=1+B_0^2r^2\sin^2\theta$ then the metric of the magnetized solution takes the simple form:
\beqs
ds^2&=&\Lambda^2\bigg[-f(r)dt^2+\frac{dr^2}{1-\frac{2m(r)}{r}}+r^2d\theta^2\bigg]+\frac{r^2\sin^2\theta}{\Lambda^2}d\varphi^2,
\eeqs
where the ansatz for the electromagnetic potential is $A_{\varphi}=\frac{B_0r^2\sin^2\theta}{\Lambda}$. Here $B_0$ is a constant parameter, which can be related to the value of the magnetic field on the axis.

The Einstein equations will be satisfied if the density of the fluid is $\rho=\frac{\rho^0}{\Lambda^2}$, while $p_{\theta}=\frac{p^0_t}{\Lambda^2}$ and $p_{\varphi}=\frac{p^0_t}{\Lambda^2}-\frac{2\rho B_0^2r^2\sin^2\theta}{\Lambda}$. Finally, the Maxwell equations are satisfied if the $4$-current is $J^{\mu}=\frac{2\rho B_0r^2\sin^2\theta}{\Lambda^2}\delta^{\mu}_{\varphi}$.
However, a study of the properties of this new exact magnetized solution will be left for further work. 

Another interesting extension of the present work would involve a study of the influence of the charged black hole with the dark halo on the propagation of various fields in this background, on the lines of the studies presented in \cite{Dariescu:2017ima}, \cite{Dariescu:2018dyy}, \cite{Dariescu:2010zz}.  

Work on these matters is in progress and it will be presented elsewhere. 

\vspace{10pt}


\end{document}